# Searching inhibitors for three important proteins of COVID-19 through molecular docking studies

Seshu Vardhan and Suban K Sahoo*

*Department of Applied Chemistry, S.V. National Institute of Technology (SVNIT), Surat-395007, India. (E-mail: sks@chem.svnit.ac.in, suban_sahoo@rediffmail.com)*

## Abstract

The lack of recommended drugs or vaccines to deal with the COVID-19 is the main concern of this pandemic. The approved drugs for similar health problems, drugs under clinical trials, and molecules from medicinal plants extracts are investigated randomly to deal with the COVID-19 infection. Molecular docking, one of the best approach to search therapeutically potent drugs/molecules in real time with possible hope to apply on COVID-19. In this communication, molecular docking studies of 18 ligands were carried out with the three therapeutic target proteins of SARS-CoV-2, i.e., RNA-dependent RNA polymerase (RdRp), angiotensin-converting enzyme 2 (ACE2) and spike glycoprotein (SGp). The obtained results revealed that the phytochemicals showed better dock score in compared to the drugs paracetmol and hydroxychloroquine. Combining the dock score and medicinal properties, we believe the terpenoids based phytochemicals limonin and scopadulcic acid B can be further explored for potential use against COVID-19.

## 1. Introduction

The virus strain severe acute respiratory syndrome coronavirus 2 (SARS-CoV-2) is a single-stranded RNA virus caused the novel coronavirus disease COVID-19, a respiratory malady with the common symptoms of shortness of breath, cough and fever [1-4]. This disease also reported to show other symptoms like muscle aches, loss of smell, fatigue and abdominal pain [5]. The first infected patient was detected in December, 2019 at Wuhan, China [6]. Afterwards, total 12768307 confirmed COVID-19 cases were reported globally with the death of 566654 by 13$^{th}$ July 2020. The outbreak of SARS-CoV-2 was declared as a public health emergency of international concern (PHEIC) and a pandemic respectively on 30$^{th}$ Jan 2020 and 11$^{th}$ March 2020 by World Health Organization (WHO).

The virus generally spread from the infected person through close contact along with the droplets spilled during talking, coughing and sneezing [7]. After infection with the virus, the symptoms likely to appear within two to fourteen days that depends on the age of person and weak immunity due to other illness like diabetes, asthma, heart ailment etc. [8]. The lack of recommended drugs or vaccines to deal with the COVID-19 along with the human to human transmission nature is the main concern of this pandemic. Therefore, at present scenario, efforts have been made to identify the infected persons through rapid diagnosis followed by quarantined them to stop the further spread of this disease. Also, other recommended steps such as using masks, washing hands with soap and maintaining social distancing are suggested to control the spread of this virus. Simultaneously, the approved drugs, drugs under clinical trial and molecules from medicinal plants extracts are investigated randomly to deal with the COVID-19 infection [9]. In searching drugs/molecules from a library that contained in lakhs, the computational approaches like molecular docking, drug-likeness screening, simulations etc. can expedite the research on drug discovery for COVID-19 [10].

In this communication, the molecular docking studies of sixteen phytochemicals and two FDA-approved drugs (paracetamol and hydroxychloroquine) were performed with the three therapeutic target proteins of SARS-CoV-2, i.e., RNA-dependent RNA polymerase (RdRp), angiotensin-converting enzyme 2 (ACE2) and spike glycoprotein (SGp). The important proteins associated with the spherical-shaped SARS-CoV-2 virus and the interaction with the human host cell is shown in **Fig. 1**. The envelope of the virus consists of a lipid bilayer made of membrane, envelope and spike structural proteins [11-13]. The club-shaped spike glycoprotein initiates the entry of the virions into host cells by interacting with human ACE2 and TMPRSS2 host cells. Once the virus entered host cells, RdRp protein play pivotal role in the replication and transcription of SARS-CoV-2. Therefore, the proteins RdRp, ACE2, and SGp were chosen as therapeutic target proteins to search inhibitors for COVID-19. The sixteen phytochemicals were selected from various classes like triterpenoids, diterpenoids, flavonoids, polyphenols, alkaloids etc. with the known diverse biological activities like antiviral, anticancer, antioxidant agent, anti-inflammatory agent, anti-dengue, antibacterial and antimicrobial. The FDA-approved drugs hydroxychloroquine and paracetamol were used as control for comparison, as they are claimed to be effective against SARS-CoV-2 [14-16].

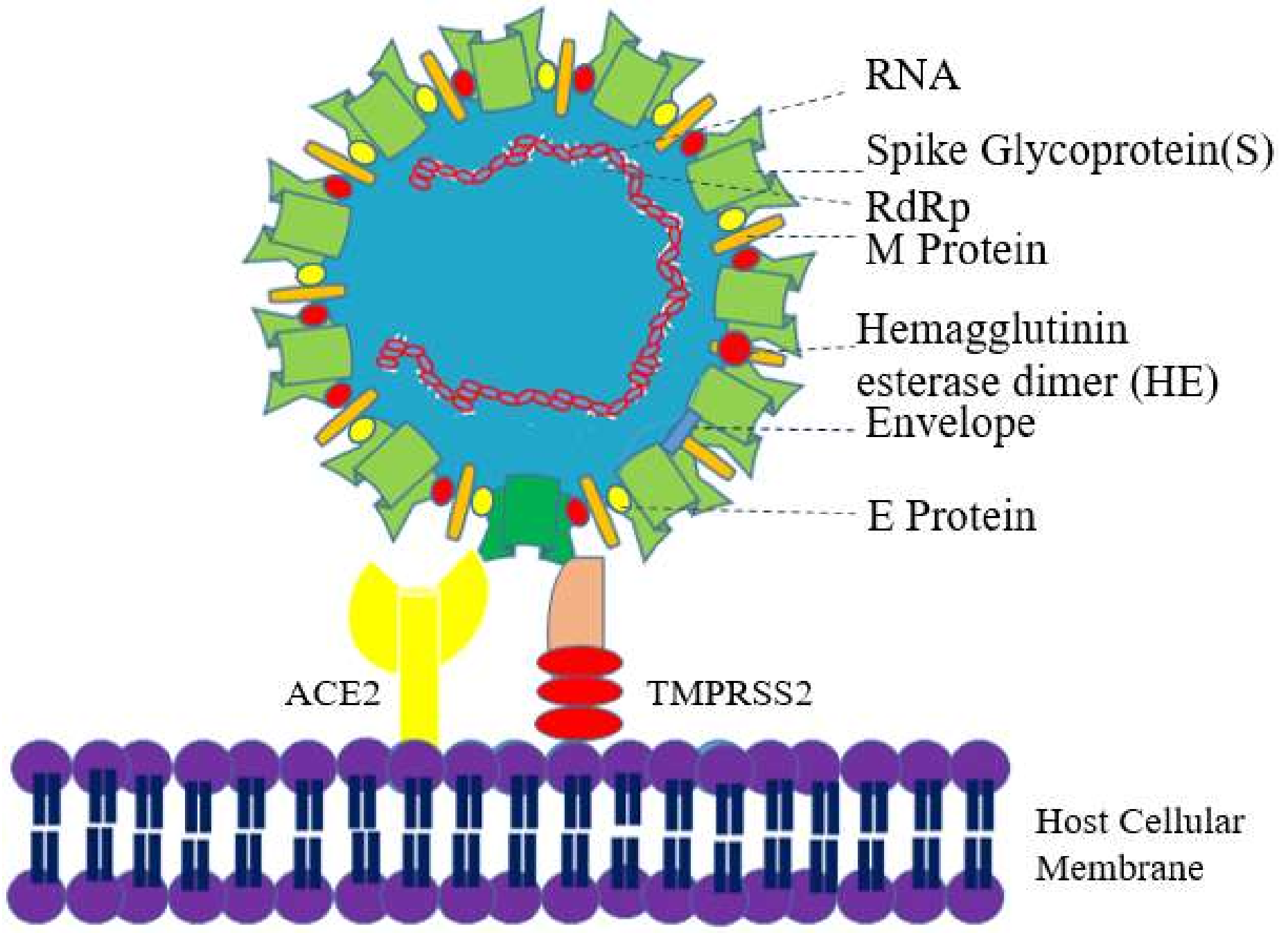

**Fig. 1.** The structure and interaction of SARS-CoV-2 virus with the host cells ACE2 and TMPRSS2.

## 2. Experimental

2.1. Preparation of ligands and protein structures for docking

To study the protein-ligand interactions, the ligands in 'SDF' format were retrieved directly from the PubChem (National Library of Medicine) [17] and converted to 'PDB' format. The crystallographic 3D structures in complex with cofactors of RdRp (PDB ID: 6M71) [17], SGp (PDB ID: 2GHV) [19] and ACE2 (PDB ID: 6M1D) [20] were collected from RSCB protein data bank (http://www.rscb.org). The proteins were modelled using the protein template of PDB structure from Swiss model online server, and were validated using (https://servicesn.mbi.ucla.edu/PROCHECK/) authenticated online server. Then, the 3D protein structures were analysed for torsion angles and missing residues using Biovia Drug Discovery Tool by plotting Ramchandram plot.

2.2. Procedure for molecular docking

All molecular docking were done by using Autodock 4.2 to examine the protein-ligand interactions at the active site and to estimate the binding affinity scores [21]. The preparation of proteins and ligands for molecular docking were done by using the MGLTools. The water molecules were removed, added hydrogen's, confirmed torsion angles and added Kollman charges. The grid box was generated to localise the binding positions for site specific docking. The modelled RdRp protein grid dimensions (Å) generated for site specific docking at x: 45.6255, y: 61.9912, z: 47.4754. The grid dimensions (x: 80.9336, y: 55.4077, z: 87.9166) and (x: 61.1524, y: 65.2399, z: 56.5904) were generated respectively for ACE2 and SGp to cover the active site. The Lamarckian Genetic Algorithm (LA) protocol was applied to perform molecular docking. The polar contacts, Van der Waals forces and other non-covalent interactions between the protein and the docked ligand were predicted using the autodock computational method. The output files generated were converted to protein-ligand complex, and analysed using BioVia Discovery Studio Software.

## 3. Results and discussion

The binding energies of the sixteen phytochemicals and the two FDA-approved drugs with the therapeutic target proteins of SARS-CoV-2, i.e., RdRp, ACE2, and SGp are summarized in **Table 1**. The molecular docking results revealed that the limonin and scopadulcic acid B are the top ranked phytochemicals than the other examined phytochemicals, and the FDA-approved drugs hydroxychloroquine and paracetamol used as a control for comparison [16]. Examining all the phytochemicals, the binding affinity scores revealed the binding order limonoid>diterpenoid>polyphenol≈flavonoid≈alkaloid. The phytochemical limonin showed highest dock score with the proteins ACE2 and RdRp, whereas the scopadulcic acid B with SGp. Limonin is the first isolated tetranortriterpenoid that cause bitterness in citrus and one of the limonoid known for a wide range of biological activities like antimalarial, anticancer, antiviral, antibacterial, antifungal along with other pharmacological activities on humans [22]. Similar to limonin, scopadulcic acid B, a diterpenoid is also known for diverse biological activities like antiviral, antimalarial, antitumor, inhibition of bone resorption and gastric acid secretion [23]. Therefore, the protein-ligand interactions of these two terpenoids based phytochemicals (limonin and scopadulcic acid B) were examined in details with the therapeutic target proteins of SARS-CoV-2.

**Table 1.** Comparative dock score of the ligands with the therapeutic target proteins of COVID-19.

| **Classifications** | **Ligands** | **Mol. Wt. (g/mol)** | **ACE2** | **RdRp** | **SGp** |
|---|---|---|---|---|---|
| Limonoid | Limonin | 470.5 | **-8.9** | **-9.0** | **-8.4** |
| Polyphenol | Ellagic acid | 302.19 | -8.4 | -8.1 | -7.5 |
| Flavonoid | Baicalein | 446.4 | -8.3 | -8.1 | -7.6 |
| Diterpenoid | Scopadulcic acid B | 438.6 | **-8.2** | **-8.6** | **-8.8** |
| Limonoid | Nimbolide | 466.5 | -8.0 | -7.6 | -7.9 |
| Triterpenoid | Dammarenolic acid | 458.7 | -7.9 | -7.2 | -6.7 |
| Flavonoid | Quercetin | 302.23 | -7.9 | -7.3 | -7.1 |
| Methylated phenol | Tocopherol | 430.7 | -7.8 | -5.5 | -6.0 |
| Phenylpropo noid | 1,5-Dicaffeoylquinic acid | 516.4 | -7.6 | -6.9 | -7.0 |
| Flavonoid | Kaempferol | 286.24 | -7.6 | -7.4 | -7.2 |
| Flavonoid | 5,7,4’-Trihydroxy-8-methoxy flavone | 302.26 | -7.4 | -7.1 | -7.0 |
| Alkaloid | Piperine | 285.34 | -7.1 | -7.4 | -7.2 |
| Flavonoid | Chalcone | 208.25 | -6.9 | -6.6 | -6.1 |
| Vitamin $A_1$ | Retinol | 286.5 | -6.6 | -6.8 | -7.2 |
| Flavonoid | Tangeretin | 372.4 | -6.4 | -6.9 | -6.4 |
| Phenolic acid | Gallic Acid | 170.12 | -6.4 | -5.9 | -5.7 |
| Drug | Acetaminophen(Paracetmol) | 151.16 | -6.5 | -7.8 | -8.2 |
| Drug | Hydroxychloroquine | 335.9 | -6.2 | -6.0 | -5.8 |

The RdRp catalyses the RNA replication from template RNA strand. The modelled protein structure of SARS CoV-2 RdRp comprises the nsp12 bound to nsp7 and nsp8 domains. The nsp12 domain involved in RNA polymerase activity that contains 398-919 amino acid residues, in which the catalytic residues were localised [24]. RdRp hydrophobic cavities at active N-terminal and C-terminals involved in catalysis of RNA polymerization. The molecular docking of RdRp with the phytochemicals resulted efficient binding of limonin at β-sheet nsp12 polymerase residues (**Fig. 2**). Limonin with RdRp showed binding energy of -9.0 kcal/mol, and binds firmly with the amino acid residues in the RdRp tunnel can possibly obstruct the binding of RNA template. In addition, the limonin showed comparable binding affinity like the drug remdesivir whereas better potency than the drugs umifenovir and favipiravir at the active site of RdRp [25]. Limonin interacts with the active residue VAL557 by van der Waals (VDW) interaction. Limonin formed hydrogen bonds with the residues THR556, SER682 and LYS621 that maintains the strong affinity with the target protein. Limonin posing π-alkyl interactions at the active site residues ARG553, TYR455, ASP623, and VDW interactions to residues SER759, ARG624, ASP452, TYR619, GLY616, ASP618. In compared to limonin, other phytochemicals known for antioxidant and antiviral properties like scopadulcic acid B, ellagic acid and baicalein showed binding affinity to RdRp active site with binding energy of -8.6, -8.1 and -8.1 kcal/mol, respectively.

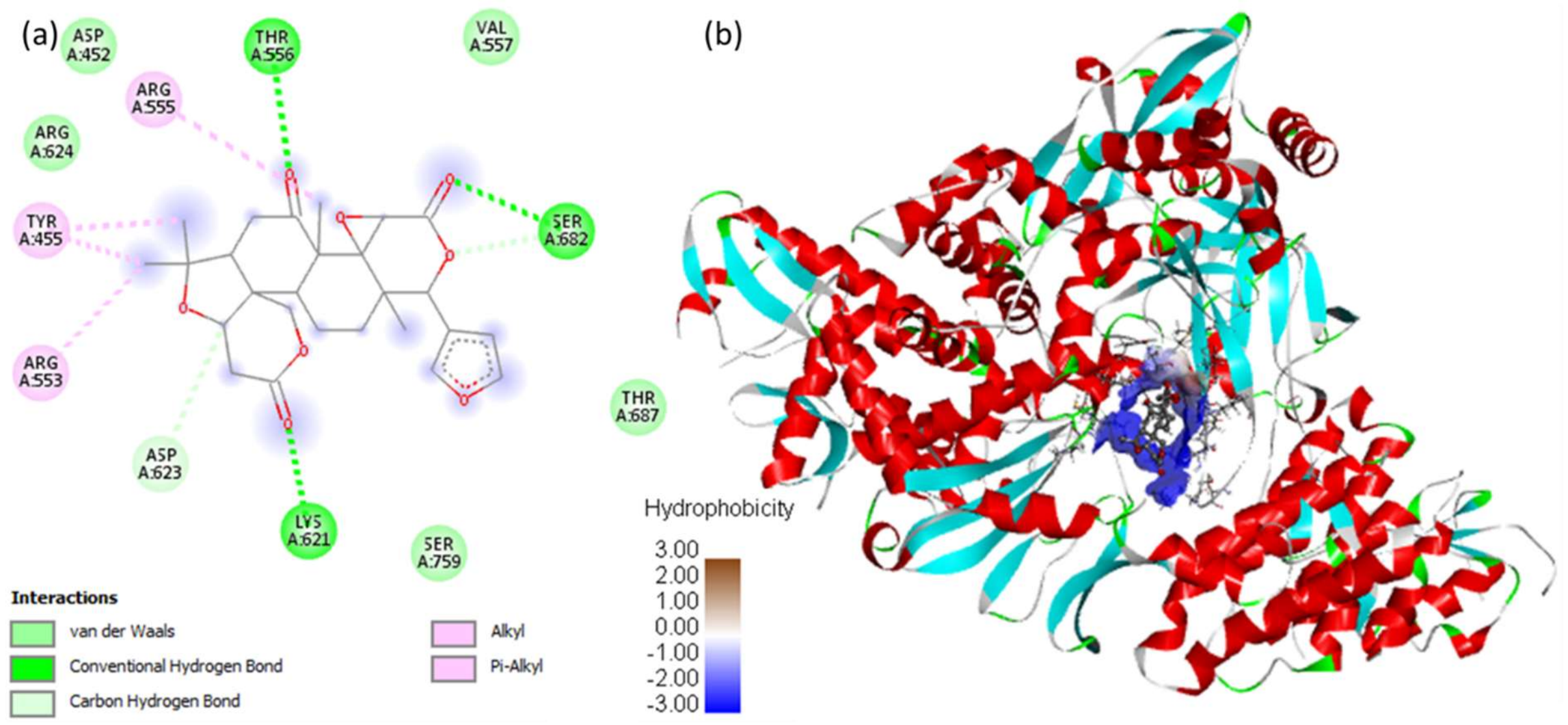

**Fig. 2.** (a) 2D animated pose showing non-covalent interactions between limonin and RdRp, and (b) 3D representation showing the position of limonin within the cavity of RdRp.

Homo sapiens angiotensin-converting enzyme 2 (ACE2) is a membrane protein that facilitates the binding to chain B domain with SGp of SARS-CoV-2 virus. The arginin and histidine residues at the active site of ACE2 plays catalytic role for substrate binding [26]. The top ranked limonin binds to the chain B residues ARG393 and HIS401 via carbon hydrogen bond and π-alkyl bond that are very close to the active site, where the binding of SGp-RBD takes place (**Fig. 3**). The non-covalent interactions are abundant due to multiple rings with oxygen-atoms in limonin that are interacting with TRP69, LEU73, PHE40, THR347, TRP349, ALA348, ASP382, ASN394, PHE390 and LEU391. Limonin shows the binding energy of -8.9 kcal/mol and actively targeting the ACE2 at cellular surface binding domain with π-alkyl interactions to TRP203, and VDW interactions with GLU398, SER511, ASP206, GLN102, GLY205, TYR202, TYR199, LYS187, ASP509, TYR510. Other phenolic, flavonoid and terpenoid based compounds like ellagic acid, baicalein and scopadulcic acid B are binding to the ACE2 active site with the respective binding energy of -8.4, -8.3 and -8.2 kcal/mol.

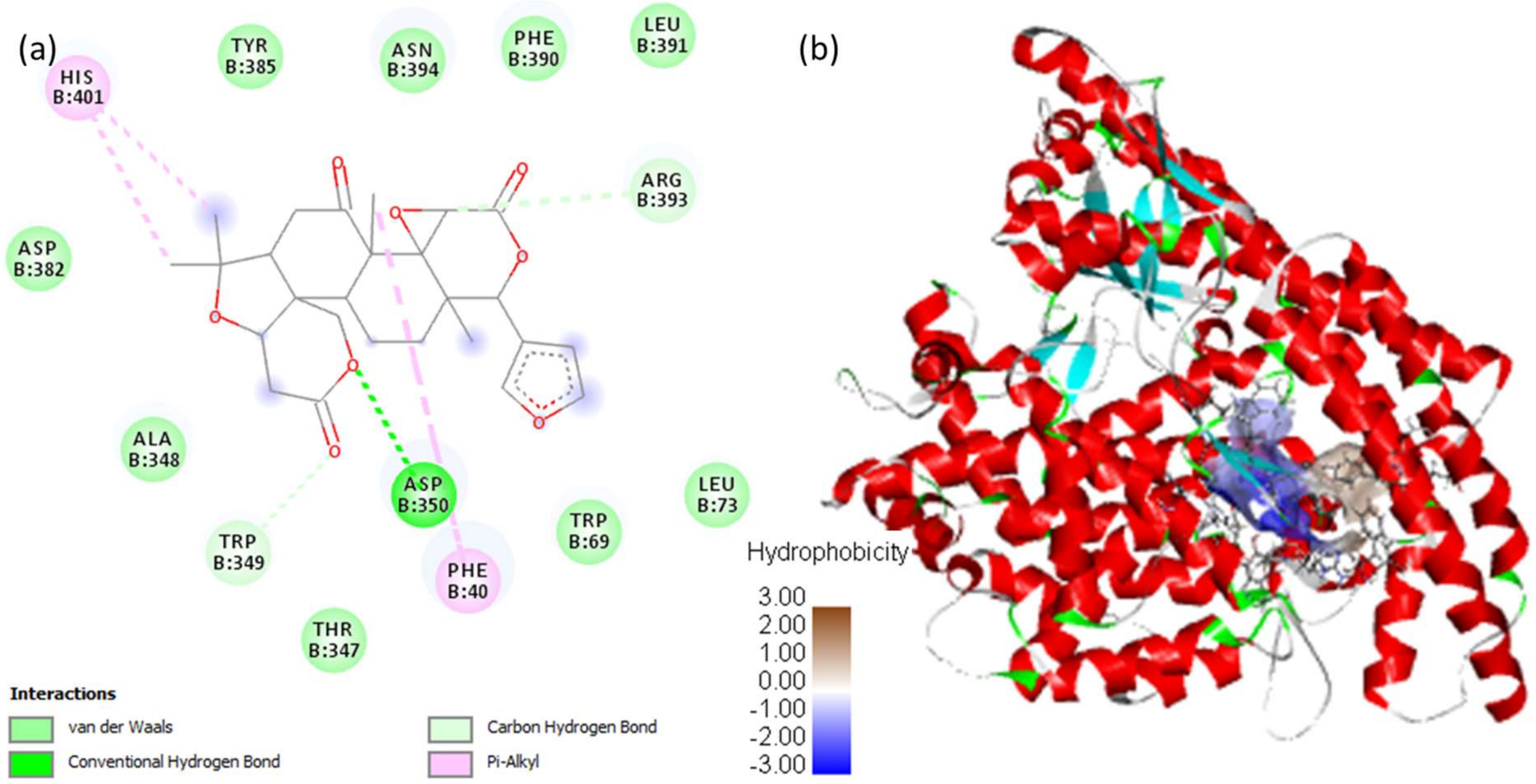

**Fig. 3.** (a) 2D animated pose showing non-covalent interactions between liminon and ACE2, (b) 3D representation showing the position of limonin within the cavity of ACE2.

The receptor binding domain (RBD) of spike glycoprotein (SGp) play the privotal role in binding with host receptor ACE2 to activate the virus-host interaction (**Fig. 1**). SGp-RDB helps in attaching the virus to the cell membrane of the human receptor ACE2 [27]. This also internalizes the virus into the endosomes where the conformational changes take place in the spike glycoprotein. Scopadulcic acid B, the diterpenoid based phytochemical pose strong binding affinity of -8.8 kcal/mol to SGp chain C residues that forming hydrogen bonds to ASN424, ASN427 and ARG426 residues, and VDW interactions with ILE428, SER362, PHE360, TRP423, SER358, THR359, THR425 and ILE489 (**Fig. 4**). These interactions may collectively hindered the SGp-RBD docking to host receptor ACE2. Limonin, with comparable binding at the active site is docked at the hydrophobic cavities of SGp-RBD chain C and E residues VAL394, ARG395, GLN401, TYR494, GLY391, ILE489, SER362 with the dock score of -8.4 kcal/mol. In addition, the binding affinity of scopadulcic acid B and limonin at the active site of spike glycoprotein is comparably better than the drug remdesivir [28]. With SGp-RBD, other phytochemicals like nimbolide, baicalein and ellagic acid poses binding affinity to active site with the binding energy of -7.9, -7.6 and -7.5 kcal/mol, respectively.

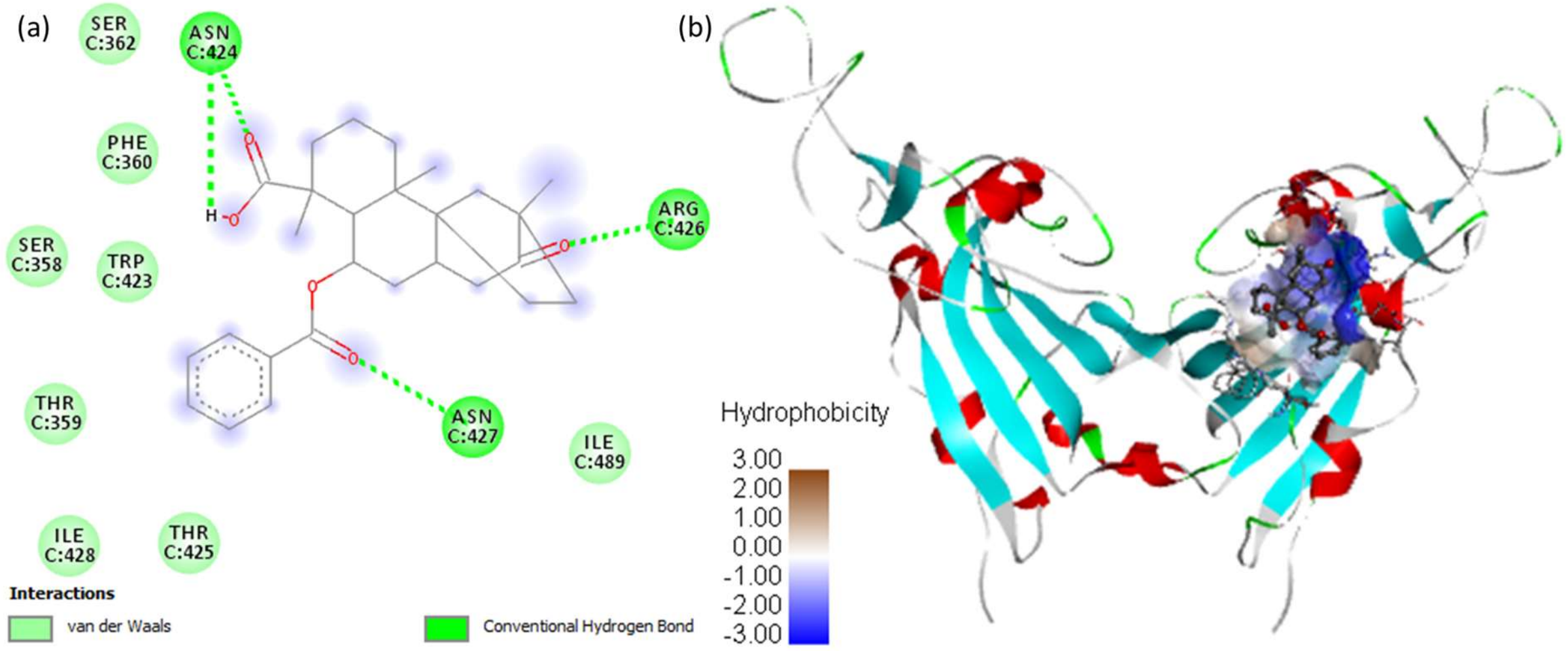

**Fig. 4.** (a) 2D animated pose showing non-covalent interactions between scopadulcic acid B and SGp, (b) 3D representation showing the position of scopadulcic acid B within the cavity of SGp.

## 4. Conclusions

In summary, the site specific molecular docking studies were performed by selecting sixteen medicinally important phytochemicals from various families like limonoids, triterpenoids, diterpenoids, flavonoids, alkaloids and polyphenols etc. with three therapeutic target proteins (RdRp, ACE2, and SGp) of SARS-CoV-2, and compared with the two FDA-approved drugs. The binding affinity scores revealed that the limonin and scopadulcic acid B showed the higher dock score than the other examined phytochemicals and the approved drugs hydroxychloroquine and paracetamol. Limonin, the tetranortriterpenoid found in citrus fruits with bitter taste, known to inhibit the replication of retroviruses like HTLV-I and HIV-1, showed the higher binding affinity score towards RdRp and ACE2. The diterpenoid scopadulcic acid B, known for good activity against the herpes simplex virus (HSV-1), showed the higher dock score with the spike glycoprotein followed by limonin. Overall, based on the binding order of examined phytochemicals (limonoid>diterpenoid>polyphenol≈flavonoid≈alkaloid) at the active site of the target proteins of SARS-CoV-2 and the abundance of the top ranked terpenoids based phytochemicals in medicinal plants like tulsi, neem, licorice, citrus etc., the outcomes of this work can be used for searching appropriate therapeutic approach for COVID-19. Finally, despite a small library taken for virtual screening, we proposed the terpenoids based phytochemicals like limonin and scopadulcic acid B can be examined further for use against the COVID-19.

## Declaration of competing interest

The authors declare that they have no known competing financial interests or personal relationships that could have appeared to influence the work reported in this paper.

## Acknowledgments

Authors are thankful to Direct, SVNIT for providing necessary research facilities and infrastructure.